\documentclass[10pt, conference, letterpaper]{IEEEtran}
\usepackage[letterpaper, left=0.65in, right=0.65in, bottom=1.01in, top=0.75in]{geometry}
\IEEEoverridecommandlockouts
\usepackage{cite}
\usepackage{amsmath,amssymb,amsfonts}
\usepackage{graphicx}
\usepackage[caption=false,font=footnotesize]{subfig}
\usepackage{textcomp}
\usepackage{xcolor}
\usepackage{tabularx}
\usepackage{multirow}
\usepackage{diagbox}
\usepackage[binary-units]{siunitx} 
\DeclareSIUnit{\bps}{bps}
\usepackage{flushend}
\usepackage{amsthm}
\def \tr {\mathrm{tr}}

\begin{document}

\title{Heterogeneous Massive MIMO: A Cost-Efficient Technique for Uniform Service in Cellular Networks}

\author{\IEEEauthorblockN{Wei Jiang\IEEEauthorrefmark{1} and Hans Dieter Schotten\IEEEauthorrefmark{2}}
\IEEEauthorblockA{\IEEEauthorrefmark{1}German Research Center for Artificial Intelligence (DFKI)\\Trippstadter Street 122,  Kaiserslautern, 67663 Germany\\
  }
\IEEEauthorblockA{\IEEEauthorrefmark{2}Technische Universit\"at  (RPTU) Kaiserslautern\\Building 11, Paul-Ehrlich Street, Kaiserslautern, 67663 Germany\\
 }
}
\maketitle

\begin{abstract}
Massive multi-input multi-output (MIMO) has evolved along two tracks: cellular and cell-free, each with unique advantages and limitations. The cellular approach suffers from worse user spectral efficiency at cell edges, whereas the cell-free approach incurs high implementation costs due to a large-scale distributed infrastructure. This paper introduces a novel networking paradigm, termed heterogeneous massive MIMO (HmMIMO), which seamlessly integrates co-located and distributed antennas. Differing from two conventional paradigms, HmMIMO remains a base station with a large antenna array at the center of each cell, aided by distributed antennas deployed at cell edges. Our findings demonstrate that this paradigm achieves a favorable trade-off between performance and implementation complexity.
\end{abstract}

\section{Introduction}

In cellular networks, a base station (BS) is deployed at the center of each cell within a network of cells. Users located near the BS generally experience high quality of service (QoS). However, many users at the cell edge suffer from low QoS due to weaker signal strength, resulting from significant distance-dependent path loss, strong inter-cell interference, and frequent handover issues inherent to the cellular structure \cite{Ref_jiang2024TextBook}. Unfortunately, the performance gap between the cell center and edge is not trivial; in fact, it is substantial, with differences reaching up to hundreds of times \cite{Ref_non2017minimum}.

Cell-free massive MIMO (CFmMIMO) has recently drawn significant attention in both academia and industry \cite{Ref_jiang20236G_ch09}. It eliminates the layout of cells, using a multitude of distributed access points (APs) that cooperatively serve a relatively smaller number of users \cite{Ref_nayebi2017precoding}. By providing consistent QoS for all users, CFmMIMO effectively addresses the challenge of under-served areas at the edges of traditional cellular networks.
Despite its considerable potential, CFmMIMO is still hard to implement, where connecting hundreds of distributed APs through a fronthaul network is not only costly but sometimes infeasible. Deploying a conventional wireless network is already a demanding task, given the difficulty of acquiring and maintaining wireless sites. The challenge is intensified in a CFmMIMO network, as it requires identifying hundreds of appropriate sites within a small area for AP installation and deploying a massive-scale fiber-cable network to connect each AP \cite{Ref_jiang2024cost}. 

To lower the complexity of implementation, this paper proposes a novel paradigm for cellular networking, termed heterogeneous massive MIMO (HmMIMO), which seamlessly integrates co-located and distributed antennas. A \textit{central base station (cBS)} with a large antenna array is positioned at the center of each cell, linked to multiple \textit{edge access points (eAPs)} distributed at the cell edges and in dead spots. The cBS is the main signal transceiver of the cell while also acting as the central processing unit (CPU) for the eAPs. Intuitively, this heterogeneous paradigm can reduce the implementation costs of the distributed network, as the large number of co-located antennas at the cBS do not require individual site acquisition or fiber connections. Additionally, our numerical results demonstrate that HmMIMO offer uniformly high-quality service by maintaining a comparable worst-case per-user rate as conventional CFmMIMO systems.

The structure of the paper is as follows. The next section introduces the system model. Section III analyzes the achievable spectral efficiency (SE) of uplink (UL), while Section IV addresses downlink (DL) transmission. Section V presents numerical results, and conclusions are drawn in Section VI.

\section{System Model}

\begin{figure}[!t]
    \centering
    \includegraphics[width=0.45\textwidth]{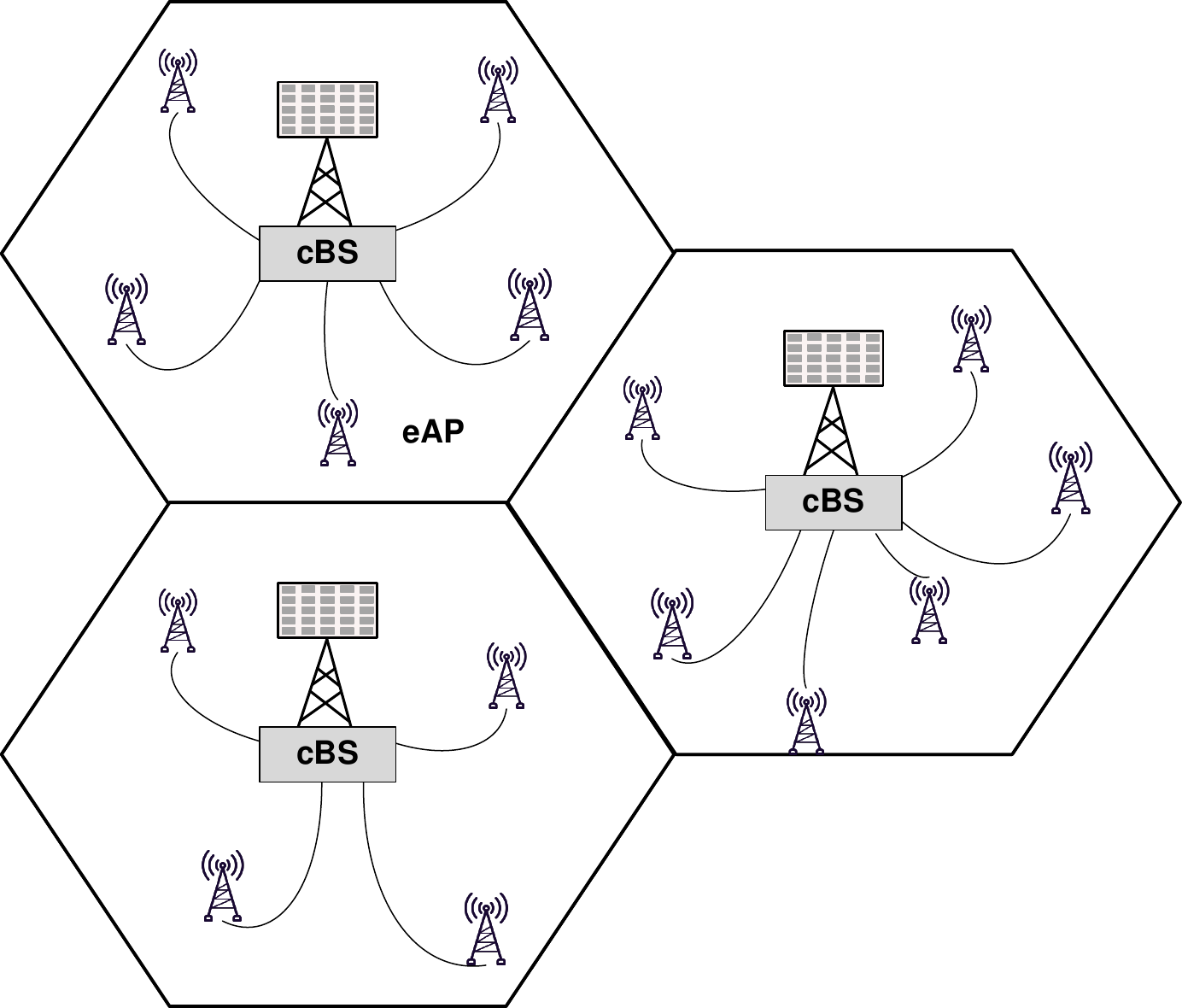} 
    \caption{Schematic diagram of a HmMIMO network. }
    \label{fig:heteromimo}
\end{figure}

This paper introduces a novel cellular networking paradigm, as shown in \figurename~\ref{fig:heteromimo}. Like the conventional cellular massive MIMO (CmMIMO) \cite{Ref_marzetta2015massive}, the coverage area comprises multiple cells. Each cell has a cBS at the cell center and multiple eAPs. The eAPs enhance service quality in areas with weak cBS signals, such as cell-edge regions and dead spots \cite{Ref_jiang2023celledge}. Acting as both primary transceiver and central processor, the cBS coordinates all eAPs within its cell. Compared to CFmMIMO, this heterogeneous approach lows distributed network costs by reducing the number of wireless AP sites and connectivity requirements.

To be specific, the network contains $C$ cells. A typical cell $c \in \{1,\ldots,C\}$ includes:
\begin{itemize}
    \item A cBS with $N_b$ co-located antennas
    \item $L_c$ single-antenna eAPs
    \item $K_c$ single-antenna users
\end{itemize}
This configuration aligns with prior CFmMIMO studies \cite{Ref_ngo2017cellfree,Ref_nayebi2017precoding} while maintaining the cBS's large-scale array. We adopt a block fading model where channels remain constant over $\tau_c$-length coherence blocks. Channels are characterized as:
\begin{itemize}
    \item Scalar link between UE $k\in \{1,\ldots,K_{c}\}$ (cell $c$) and eAP $l \in \{1,\ldots,L_{\iota}\}$ (cell $\iota$): $h_{ck}^{\iota l} \in \mathbb{C}$, which follows $h_{ck}^{\iota l} \sim \mathcal{CN}(0, \beta_{ck}^{\iota l})$, and $\beta_{ck}^{\iota l}$ is large-scale fading coefficient.
    \item Vector channel between cBS in cell $\iota$ and UE $k$ (cell $c$): $\mathbf{h}_{ck}^{\iota0} \in \mathbb{C}^{N_b}$, where $\mathbf{h}_{ck}^{\iota0} \sim \mathcal{CN}(\mathbf{0}, \beta_{ck}^{\iota 0}\mathbf{I}_{N_b})$.
\end{itemize}

TDD operation dividing each coherence block into: UL training ($\tau_p$ symbols), UL data transmission, and DL data transmission. During UL training, users simultaneously transmit pilot sequences. Using linear minimum mean-square error (MMSE) estimation \cite{Ref_bjornson2020making} yields channel estimates $\hat{h}_{ck}^{\iota l}$, following a complex normal distribution $\mathcal{CN}(0,\alpha_{ck}^{\iota l})$ with $\alpha_{ck}^{\iota l}=\frac{p_u(\beta_{ck}^{\iota l})^2}{p_u \beta_{ck}^{\iota l} + \sigma_n^2}$, where $p_u$ represents the power of the UE transmitter, and $\sigma_n^2$ is the variance of noise \cite{Ref_jiang2021cellfree}. The estimation error is defined as $\tilde{h}_{ck}^{\iota l} = h_{ck}^{\iota l}-\hat{h}_{ck}^{\iota l}$, following the distribution of $\mathcal{CN}(0,\beta_{ck}^{\iota l}-\alpha_{ck}^{\iota l})$. The similar results apply to the estimated channels for the cBSs.

\section{Uplink Data Transmission}

During uplink transmission, all UEs simultaneously transmit their signals. The data symbols are assumed to be independent, zero-mean random variables with unit variance, where $x_{\kappa\iota}$ from UE $\kappa$ in cell $\iota$ has power coefficient $0 \leqslant \eta_{\kappa\iota} \leqslant 1$. Therefor, the symbol vector $\mathbf{x}_{\iota} = [x_{\iota 1}, \ldots, x_{\iota K_\iota}]^T$ satisfies $\mathbb{E}[\mathbf{x}_{\iota}\mathbf{x}_{\iota}^H] = \mathbf{I}_{K_{\iota}}$. A typical single-antenna eAP $l$ in cell $c$ receives:
\begin{equation} \label{GS_uplink_RxsignalAP}
    y_{cl} = \sqrt{p_u} \sum_{\iota=1}^C \sum_{\kappa=1}^{K_{\iota}} \sqrt{\eta_{\iota\kappa}} h_{\iota\kappa}^{cl} x_{\iota\kappa} + n_{cl}
\end{equation} 
where $n_{cl} \sim \mathcal{CN}(0, \sigma^2_n)$. Concurrently, the cBS in cell $c$ observes:
\begin{equation} 
    \mathbf{y}_{c0} = \sqrt{p_u} \sum_{\iota=1}^C \sum_{\kappa=1}^{K_{\iota}} \sqrt{\eta_{\iota\kappa}} \mathbf{h}_{\iota\kappa}^{c0} x_{\iota\kappa} + \mathbf{n}_{c0}
\end{equation} 
with $\mathbf{n}_{c0} \sim \mathcal{CN}(\mathbf{0}, \sigma^2_n\mathbf{I}_{N_b})$.

eAPs forward received signals to their cBS via fronthaul links. As a result, the cBS gets an aggregated signal vector, denoted as 
\begin{equation}
    \mathbf{y}_c = [\mathbf{y}_{c0}^T, y_{c1}, \ldots, y_{cL_c}]^T,
\end{equation}
which can be also expressed by
\begin{equation} \label{eqnUPLINKmodel}
    \mathbf{y}_c = \sqrt{p_u} \sum_{\iota=1}^C \sum_{\kappa=1}^{K_{\iota}} \sqrt{\eta_{\iota\kappa}} \mathbf{h}_{\iota\kappa}^{c} x_{\iota\kappa} + \mathbf{n}_{c}
\end{equation}
where:
\begin{align}
    \mathbf{n}_{c} &= [\mathbf{n}_{c0}^T, n_{c1}, \ldots, n_{cL_c}]^T \\
    \mathbf{h}_{\iota \kappa}^{c} &= [(\mathbf{h}_{\iota \kappa}^{c0})^T, h_{\iota \kappa}^{c1}, \ldots, h_{\iota \kappa}^{cL_c}]^T
\end{align}
represents the composited noise and channel vector. 
The cBS employs the maximum-ratio combining method to detect $x_{ck}$, as in \cite{Ref_ngo2017cellfree}, by correlating the received signal $\mathbf{y}_c$ with the Hermitian of the channel estimate $\hat{\mathbf{h}}_{ck}^c$, which yields a soft estimate of
\begin{align} \label{massiveMIMO:MFsoftestimateUL} \nonumber
    \hat{x}_{ck} &= \frac{1}{\sqrt{p_u}} (\hat{\mathbf{h}}_{ck}^c)^H \mathbf{y}_c \\ \nonumber
    &=  \sum_{\iota=1}^C \sum_{\kappa=1}^{K_{\iota}} \sqrt{\eta_{\iota\kappa}} (\hat{\mathbf{h}}_{ck}^c)^H \mathbf{h}_{\iota\kappa}^{c} x_{\iota\kappa} + (\hat{\mathbf{h}}_{ck}^c)^H \frac{\mathbf{n}_{c} }{\sqrt{p_u}} \\ \nonumber
    &= \sqrt{\eta_{ck}} \|\hat{\mathbf{h}}_{ck}^c\|^2  x_{ck} + \underbrace{\sqrt{\eta_{ck}} (\hat{\mathbf{h}}_{ck}^c)^H \tilde{\mathbf{h}}_{ck}^c  x_{ck}}_{\mathcal{I}_1} \\ \nonumber
    &+ \underbrace{\sum_{\substack{\kappa=1 \\ \kappa \neq k}}^{K_c} \sqrt{\eta_{c\kappa}} (\hat{\mathbf{h}}_{ck}^c)^H \mathbf{h}_{c\kappa}^c x_{c\kappa}}_{\mathcal{I}_2} + \underbrace{\sum_{\iota \neq c} \sum_{\kappa=1}^{K_{\iota}} \sqrt{\eta_{\iota\kappa}} (\hat{\mathbf{h}}_{ck}^c)^H \mathbf{h}_{\iota\kappa}^{c} x_{\iota\kappa}}_{\mathcal{I}_3} \\
    &+ \underbrace{\frac{(\hat{\mathbf{h}}_{ck}^c)^H\mathbf{n}_c}{\sqrt{p_u}}}_{\mathcal{I}_4}
\end{align}

Therefore, the uplink spectral efficiency for UE $k$ in cell $c$ is given by the expression:
\begin{equation}
    \mathsf{SE}_{ck}^{\mathsf{UL}} = \left(1 - \frac{\tau_p}{\tau_u}\right) \mathbb{E} \left[ \log_2 \left(1 + \gamma_{ck}^{\mathsf{UL}} \right) \right]
\end{equation}
with effective signal-to-interference-plus-noise ratio (SINR):
\begin{align} \label{eQn_ULspectraleff_ck}
    \gamma_{ck}^{\mathsf{UL}} = \frac{ \eta_{ck} \| \hat{\mathbf{h}}_{ck}^c \|^4 }{ (\hat{\mathbf{h}}_{ck}^c)^H \mathbf{\Xi}_{ck} \hat{\mathbf{h}}_{ck}^c }
\end{align}
where $\mathbf{\Xi}_{ck} = \sum\limits_{\substack{\kappa=1 \\ \kappa \neq k}}^{K_c} \eta_{c\kappa}\hat{\mathbf{h}}_{c\kappa}^c (\hat{\mathbf{h}}_{c\kappa}^c)^H + \sum\limits_{\kappa=1}^{K_c} \eta_{c\kappa} \boldsymbol{\Theta}_{c\kappa}^c + \sum\limits_{\substack{\iota=1 \\ \iota \neq c}}^C \sum\limits_{\kappa=1}^{K_{\iota}} \eta_{\iota\kappa} \mathbf{R}_{\iota\kappa}^c + \frac{\sigma_n^2}{p_u}\mathbf{I}_{M_c}$.
\begin{IEEEproof}
The interference terms from $\mathcal{I}_1$ to $\mathcal{I}_4$ in \eqref{massiveMIMO:MFsoftestimateUL} are mutually uncorrelated due to the independence of data symbols/noise, namely $\mathbb{E}[x_{ck}^*x_{\iota\kappa}]=0$, for all $c\neq \iota$ or $k\neq \kappa$. This implies $ \mathbb{E}[|\sum_{i=1}^4\mathcal{I}_i|^2]=\sum_{i=1}^4 \mathbb{E}[|\mathcal{I}_i|^2]$.

Note that $\{\hat{\mathbf{h}}_{c\kappa}^c\}_{\kappa=1,\ldots,K_c}$ are \textit{deterministic} for the cBS of cell $c$, the variances of the first interference term are computed as
\begin{equation}  \label{APPEQ1} \nonumber
    \mathbb{E}\left[|\mathcal{I}_1|^2\right]  =  \eta_{ck} (\hat{\mathbf{h}}_{ck}^c)^H \mathbb{E}\left[\tilde{\mathbf{h}}_{ck}^c(\tilde{\mathbf{h}}_{ck}^c)^H\right] \hat{\mathbf{h}}_{ck}^c,
\end{equation}
where the correlation matrix of the channel estimation error \(\tilde{\mathbf{h}}_{\iota \kappa}^{c}\) is given by:
\begin{align} \nonumber
    \boldsymbol{\Theta}_{c\kappa}^c&=\mathbb{E}\left[\tilde{\mathbf{h}}_{ck}^{c} (\tilde{\mathbf{h}}_{ck}^{c})^H\right] = \\ & \begin{bmatrix}
(\beta_{ck}^{c0}-\alpha_{ck}^{c0}) \mathbf{I}_{N_b} & \mathbf{0} & 0\\
\mathbf{0} & \beta_{ck}^{c 1}-\alpha_{ck}^{c 1} &  \\
\vdots & \ddots & \vdots \\
\mathbf{0} & \ldots &\beta_{ck}^{c L_c}-\alpha_{ck}^{c L_c}
\end{bmatrix}
\end{align}
Hence, \eqref{APPEQ1} becomes
\begin{equation}   \nonumber
    \mathbb{E}\left[|\mathcal{I}_1|^2\right]  =  \eta_{ck} (\hat{\mathbf{h}}_{ck}^c)^H \boldsymbol{\Theta}_{c\kappa}^c \hat{\mathbf{h}}_{ck}^c.
    \end{equation}
And therefore, we have
\begin{align} \nonumber
    &\mathbb{E}\left[|\mathcal{I}_2|^2\right]=\sum_{{\kappa}=1,{\kappa}\neq k}^{K_c} \eta_{c\kappa}\mathbb{E}\left[\left| (\hat{\mathbf{h}}_{ck}^c)^H \mathbf{h}_{c\kappa}^c  \right|^2\right]\\ \nonumber
     &=\sum_{{\kappa}=1,{\kappa}\neq k}^{K_c} \eta_{c\kappa}(\hat{\mathbf{h}}_{ck}^c)^H \mathbb{E}\left[  \mathbf{h}_{c\kappa}^c (\mathbf{h}_{c\kappa}^c)^H \right] \hat{\mathbf{h}}_{ck}^c \\ \nonumber
    &=\sum_{{\kappa}=1,{\kappa}\neq k}^{K_c} \eta_{c\kappa}(\hat{\mathbf{h}}_{ck}^c)^H\left(\hat{\mathbf{h}}_{c\kappa}^c (\hat{\mathbf{h}}_{c\kappa}^c)^H+ \mathbb{E}\left[ \tilde{\mathbf{h}}_{c\kappa}^c (\tilde{\mathbf{h}}_{c\kappa}^c)^H \right] \right) \hat{\mathbf{h}}_{ck}^c \\
   &=\sum_{{\kappa}=1,{\kappa}\neq k}^{K_c} \eta_{c\kappa}(\hat{\mathbf{h}}_{ck}^c)^H\left(\hat{\mathbf{h}}_{c\kappa}^c (\hat{\mathbf{h}}_{c\kappa}^c)^H+ \boldsymbol{\Theta}_{c\kappa}^c \right) \hat{\mathbf{h}}_{ck}^c.   
\end{align}
In contrast, the cBS of cell $c$ does not know $\{\hat{\mathbf{h}}_{\iota\kappa}^c\}$ because it lacks direct connections to the eAPs in other cells. Consequently, the calculation of the third term is different, i.e.,
\begin{align} \nonumber
    \mathbb{E}\left[|\mathcal{I}_{3}|^2\right]&=\sum_{\iota=1,\iota\neq c}^C \sum_{\kappa=1}^{K_{\iota}} \eta_{\iota\kappa}\mathbb{E}\left[\left| (\hat{\mathbf{h}}_{ck}^c)^H \mathbf{h}_{\iota\kappa}^c  \right|^2\right]\\ \nonumber
    &=\sum_{\iota=1,\iota\neq c}^C \sum_{\kappa=1}^{K_{\iota}} \eta_{\iota\kappa}(\hat{\mathbf{h}}_{ck}^c)^H \mathbb{E}\left[  \mathbf{h}_{\iota\kappa}^c (\mathbf{h}_{\iota\kappa}^c)^H \right] \hat{\mathbf{h}}_{ck}^c \\
   &=\sum_{\iota=1,\iota\neq c}^C \sum_{\kappa=1}^{K_{\iota}} \eta_{\iota\kappa}(\hat{\mathbf{h}}_{ck}^c)^H   \mathbf{R}_{\iota\kappa}^c  \hat{\mathbf{h}}_{ck}^c . 
\end{align}
In the above equation, the correlation matrix \(\mathbb{E}\left[\mathbf{h}_{\iota k}^{c} (\mathbf{h}_{\iota k}^{c})^H\right]\) is block-diagonal:
\begin{itemize}
    \item cBS block: \(\mathbb{E}\left[\mathbf{h}_{\iota \kappa}^{c 0} (\mathbf{h}_{\iota \kappa}^{c 0})^H\right] = \beta_{\iota \kappa}^{c 0} \mathbf{I}_{N_b}\).
    \item eAP diagonal terms: \(\mathbb{E}\left[h_{\iota \kappa}^{c l} (_{\iota \kappa}^{c l})^*\right] = \beta_{\iota \kappa}^{c l}\).
\end{itemize}
Thus, we write 
\[
\mathbf{R}_{\iota\kappa}^c=\mathbb{E}\left[\mathbf{h}_{\iota k}^{c} (\mathbf{h}_{\iota k}^{c})^H\right] = \begin{bmatrix}
\beta_{\iota \kappa}^{c 0} \mathbf{I}_{N_b} & \mathbf{0} \\
\mathbf{0} & \text{diag}(\beta_{\iota \kappa}^{c 1}, \ldots, \beta_{\iota \kappa}^{c L_c}).
\end{bmatrix}
\]
Lastly, 
\begin{equation}
    \mathbb{E}\left[|\mathcal{I}_4|^2\right] = \frac{\sigma_n^2}{p_u}(\hat{\mathbf{h}}_{ck}^c)^H \mathbf{I}_{M_c}  \hat{\mathbf{h}}_{ck}^c,
\end{equation}
where $M_c=N_b+L_cN_a$ denotes the number of service antennas in cell $c$. Thus, we get \eqref{eQn_ULspectraleff_ck}.
\end{IEEEproof}

\section{Downlink Data Transmission}

Within a coherence block, UL and DL channels are reciprocal, allowing the cBS to use UL channel estimates for DL precoding. The DL data symbols for \(K_{\iota}\) users in cell \(\iota\) are denoted by \(\mathbf{u}_{\iota} = [u_{\iota 1}, \ldots, u_{\iota K_{\iota}}]^T\), with \(\mathbb{E}[\mathbf{u}_{\iota} \mathbf{u}_{\iota}^H] = \mathbf{I}_{K_{\iota}}\). Conjugate beamforming (CBF) \cite{Ref_yang2013performance} is employed to spatially multiplex these symbols. Therefore, we have
\begin{itemize}
    \item The cBS in cell $\iota$ transmits:
    \begin{equation}
        \mathbf{s}_{\iota0} = \sum_{\kappa=1}^{K_\iota} \mathbf{D}_{\iota\kappa}^{\iota0} (\hat{\mathbf{h}}_{\iota\kappa}^{\iota0})^* u_{\iota\kappa}
    \end{equation}
    \item Each eAP $l$ in cell $\iota$ transmits:
    \begin{equation} \label{eQn:compositeTxSig_MR}
        s_{\iota l} = \sum_{\kappa=1}^{K_\iota} d_{\iota\kappa}^{\iota l} (\hat{h}_{\iota\kappa}^{\iota l})^* u_{\iota\kappa}
    \end{equation}
\end{itemize}
where $\mathbf{D}_{\iota\kappa}^{\iota0}$ is a power-control matrix (cBS) and $d_{\iota\kappa}^{\iota l}$ are scalar power coefficients (eAPs).

The received signal at user $k$ in cell $c$ is:
\begin{align} \label{eQn_downlinkModel}
    y_{ck} &= \sqrt{p_d} \sum_{\iota=1}^C (\mathbf{h}_{ck}^{\iota0})^T\mathbf{s}_{\iota0} + \sqrt{p_d} \sum_{\iota=1}^C \sum_{l=1}^{L_\iota} h_{ck}^{\iota l} s_{\iota l} + n_{ck} \\
    &= \sqrt{p_d} \sum_{\iota=1}^C (\mathbf{h}_{ck}^{\iota0})^T \sum_{\kappa=1}^{K_\iota} \mathbf{D}_{\iota\kappa}^{\iota0} (\hat{\mathbf{h}}_{\iota\kappa}^{\iota0})^* u_{\iota\kappa} \nonumber \\
    &+ \sqrt{p_d} \sum_{\iota=1}^C \sum_{l=1}^{L_\iota} h_{ck}^{\iota l} \sum_{\kappa=1}^{K_\iota} d_{\iota\kappa}^{\iota l} (\hat{h}_{\iota\kappa}^{\iota l})^* u_{\iota\kappa} + n_{ck}
\end{align}
where $n_{ck} \sim \mathcal{CN}(0, \sigma_n^2)$.

Without downlink pilots, users detect signals using channel statistics. Decomposing \eqref{eQn_downlinkModel}:
\begin{align} \label{eQn_DLGeneralSig} \nonumber
    y_{ck} = &\underbrace{\left( \mathbb{E}\left[ (\mathbf{h}_{ck}^{c0})^T \mathbf{D}_{ck}^{c0} (\hat{\mathbf{h}}_{ck}^{c0})^* \right] + \sum_{l=1}^{L_c} \mathbb{E}\left[  d_{ck}^{cl} \left|\hat{h}_{ck}^{cl})\right|^2 \right] \right) u_{ck}}_{\text{desired signal}} \\
    &+ \underbrace{\left( \begin{aligned} 
        & (\mathbf{h}_{ck}^{c0})^T \mathbf{D}_{ck}^{c0} (\hat{\mathbf{h}}_{ck}^{c0})^* - \mathbb{E}\left[ (\mathbf{h}_{ck}^{c0})^T \mathbf{D}_{ck}^{c0} (\hat{\mathbf{h}}_{ck}^{c0})^* \right] \\ \nonumber
        & + \sum_{l=1}^{L_c} \left( d_{ck}^{cl} \left|\hat{h}_{ck}^{cl})\right|^2 - \mathbb{E}\left[ d_{ck}^{cl} \left|\hat{h}_{ck}^{cl})\right|^2 \right] \right)
    \end{aligned} \right) u_{ck}}_{\mathcal{J}_1} \\\nonumber
    &+ \underbrace{ (\mathbf{h}_{ck}^{c0})^T \sum_{\kappa \neq k} \mathbf{D}_{c\kappa}^{c0} (\hat{\mathbf{h}}_{c\kappa}^{c0})^* u_{c\kappa} + \sum_{l=1}^{L_c} h_{ck}^{cl} \sum_{\kappa \neq k} d_{c\kappa}^{cl} (\hat{h}_{c\kappa}^{cl})^* u_{c\kappa} }_{\mathcal{J}_2} \\ \nonumber
    &+ \underbrace{ \sum_{\iota \neq c} \left( \begin{aligned} &        
    (\mathbf{h}_{ck}^{\iota0})^T \sum_{\kappa} \mathbf{D}_{\iota\kappa}^{\iota0} (\hat{\mathbf{h}}_{\iota\kappa}^{\iota0})^* u_{\iota\kappa} + \\ & \sum_{l=1}^{L_\iota} h_{ck}^{\iota l} \sum_{\kappa} d_{\iota\kappa}^{\iota l} (\hat{h}_{\iota\kappa}^{\iota l})^* u_{\iota\kappa} \end{aligned} \right) }_{\mathcal{J}_3} + \frac{n_{ck}}{\sqrt{p_d}}
\end{align}

The downlink spectral efficiency is:
\begin{equation}
    \mathsf{SE}_{ck}^{\mathsf{DL}} = \mathbb{E} \left[ \log_2 \left(1 + \gamma_{ck}^{\mathsf{DL}} \right) \right]
\end{equation}
with SINR:
\begin{equation} \label{eQn:DLSINR}
    \gamma_{ck}^{DL} = \frac{  \left| \alpha_{ck}^{c0} \tr(\mathbf{D}_{ck}^{c0}) + \sum_{l=1}^{L_c} d_{ck}^{cl} \alpha_{ck}^{cl}  \right|^2  }{ \left( \begin{aligned}    
     &\sum \nolimits_{\iota=1}^C  \sum \nolimits_{\kappa=1}^{K_{\iota}}  \alpha_{\iota\kappa}^{\iota0}\beta_{c k}^{\iota0} \tr\left(( \mathbf{D}_{\iota \kappa}^{\iota 0})^2 \right) + \\
     &
     \sum \nolimits_{\iota=1}^C  \sum \nolimits_{l=1}^{L_\iota} \sum \nolimits_{\kappa=1}^{K_{\iota}} (d_{\iota\kappa}^{\iota l})^2 \beta_{ck}^{\iota l} \alpha_{\iota\kappa}^{\iota l}+ \frac{\sigma_n^2}{p_d}. \end{aligned} \right)
      } 
\end{equation}

\begin{IEEEproof}
First, we compute the expected values for both cBS and eAP components:

1. {cBS component:}
\begin{align} \label{eQn:expectedValuedforhkzero} \nonumber
    \mathbb{E} \left[ (\mathbf{h}_{ck}^{c0})^T \mathbf{D}_{ck}^{c0} (\hat{\mathbf{h}}_{ck}^{c0})^* \right] 
    &= \mathbb{E} \left[ (\hat{\mathbf{h}}_{ck}^{c0} + \tilde{\mathbf{h}}_{ck}^{c0})^T \mathbf{D}_{ck}^{c0} (\hat{\mathbf{h}}_{ck}^{c0})^* \right] \\
    &= \mathbb{E} \left[ (\hat{\mathbf{h}}_{ck}^{c0})^T \mathbf{D}_{ck}^{c0} (\hat{\mathbf{h}}_{ck}^{c0})^* \right]
\end{align}
since $\mathbb{E}[(\hat{\mathbf{h}}_{ck}^{c0})^H\tilde{\mathbf{h}}_{ck}^{c0}] = 0$. With channel estimate covariance $\mathbb{E}[\hat{\mathbf{h}}_{ck}^{c0} (\hat{\mathbf{h}}_{ck}^{c0})^H] = \alpha_{ck}^{c0} \mathbf{I}_{N_b}$, \eqref{eQn:expectedValuedforhkzero} is further derived as 
\begin{align} \label{Proof_expectedValueHk0} \nonumber
    \mathbb{E} \left[ (\hat{\mathbf{h}}_{ck}^{c0})^T \mathbf{D}_{ck}^{c0} (\hat{\mathbf{h}}_{ck}^{c0})^* \right] 
    &= \tr \left( \mathbf{D}_{ck}^{c0} \mathbb{E}[\hat{\mathbf{h}}_{ck}^{c0} (\hat{\mathbf{h}}_{ck}^{c0})^H] \right) \\ 
    &= \alpha_{ck}^{c0} \tr(\mathbf{D}_{ck}^{c0}).
\end{align}

2. {eAP components:}
\begin{align} \label{eQn_drivationyyyy_eAP}
    \mathbb{E} \left[ d_{ck}^{cl} |\hat{h}_{ck}^{cl}|^2 \right] 
    &= d_{ck}^{cl} \mathbb{E}[|\hat{h}_{ck}^{cl}|^2] 
    = d_{ck}^{cl} \alpha_{ck}^{cl}
\end{align}
since $\hat{h}_{ck}^{cl} \sim \mathcal{CN}(0, \alpha_{ck}^{cl})$.

Due to data symbol independence ($\mathbb{E}[u_{ck}^*u_{\iota\kappa}]=0$ for $c\neq \iota$ or $k\neq \kappa$), interference terms are uncorrelated:
\begin{equation}
    \mathbb{E}\left[|\mathcal{J}_1 + \mathcal{J}_2 + \mathcal{J}_3|^2\right] = \mathbb{E}\left[|\mathcal{J}_1|^2\right] + \mathbb{E}\left[|\mathcal{J}_2|^2\right] + \mathbb{E}\left[|\mathcal{J}_3|^2\right]
\end{equation}

{Variance of $\mathcal{J}_1$:} 
\begin{align} \label{eQn_Jzero} \nonumber
    \mathbb{E}\left[|\mathcal{J}_1|^2\right] &= \sum_{l=0}^{L_c} \mathbb{E} \left[ \left| T_l - \mathbb{E}[T_l] \right|^2 \right] \\
    &= \sum_{l=0}^{L_c} \left( \mathbb{E}[|T_l|^2] - \left|\mathbb{E}[T_l]\right|^2 \right)
\end{align}
where $T_0 = (\mathbf{h}_{ck}^{c0})^T \mathbf{D}_{ck}^{c0} (\hat{\mathbf{h}}_{ck}^{c0})^*$ and $T_l = d_{ck}^{cl} |\hat{h}_{ck}^{cl}|^2$ for $l \geq 1$.
For $l=0$ (cBS), we derive
\begin{align} \label{eQn_vec_expectation}
    \mathbb{E}[|T_0|^2] &= \alpha_{ck}^{c0}\beta_{ck}^{c0} \tr\left(( \mathbf{D}_{ck}^{c0})^2 \right)
\end{align}
For $l \geq 1$ (eAPs), since $\hat{h}_{ck}^{cl}$ is scalar:
\begin{align} \label{eQn_scalar_expectation}
    \mathbb{E}[|T_l|^2]  \nonumber
    &= (d_{ck}^{cl})^2 \mathbb{E}\left[ |\hat{h}_{ck}^{cl}|^4 \right] \\
    &= (d_{ck}^{cl})^2 \cdot 2(\alpha_{ck}^{cl})^2 \quad \text{(exponential RV property)}
\end{align}
Combining with \eqref{eQn_vec_expectation} and \eqref{eQn_scalar_expectation}:
\begin{align} \label{eQn_JzeroFinal}
    \mathbb{E}\left[|\mathcal{J}_1|^2\right]    
    &= \alpha_{ck}^{c0}\beta_{ck}^{c0} \tr\left( (\mathbf{D}_{ck}^{c0})^2 \right)+\sum_{l=1}^{L_c} 2(d_{ck}^{cl}\alpha_{ck}^{cl})^2
\end{align}

Variance of $\mathcal{J}_2$:
\begin{align} \label{eQn_Jtwo}
    \mathbb{E}\left[|\mathcal{J}_2|^2\right]  \nonumber
    &= \underbrace{\sum_{\kappa \neq k} \mathbb{E}\left[ \left| (\mathbf{h}_{ck}^{c0})^T \mathbf{D}_{c\kappa}^{c0} (\hat{\mathbf{h}}_{c\kappa}^{c0})^* \right|^2 \right]}_{\text{cBS terms}} \\
    &+ \underbrace{\sum_{l=1}^{L_c} \sum_{\kappa \neq k} \mathbb{E}\left[ \left| h_{ck}^{cl} d_{c\kappa}^{cl} (\hat{h}_{c\kappa}^{cl})^* \right|^2 \right]}_{\text{eAP terms}}
\end{align}

cBS terms ($l=0$) follow vector derivation:
\begin{align} \nonumber
    & \alpha_{c\kappa}^{c0}\beta_{ck}^{c0} \tr\left(( \mathbf{D}_{c\kappa}^{c0})^2 \right)
\end{align}

eAP terms ($l \geq 1$) use scalar properties:
\begin{align} \nonumber
    & \mathbb{E}\left[ \left| h_{ck}^{cl} d_{c\kappa}^{cl} (\hat{h}_{c\kappa}^{cl})^* \right|^2 \right] \\
    &= (d_{c\kappa}^{cl})^2 \mathbb{E}\left[ |h_{ck}^{cl}|^2 \right] \mathbb{E}\left[ |\hat{h}_{c\kappa}^{cl}|^2 \right] \\
    &= (d_{c\kappa}^{cl})^2 \beta_{ck}^{cl} \alpha_{c\kappa}^{cl}
\end{align}

{Variance of $\mathcal{J}_3$:} Derivation follows similar pattern with inter-cell terms.
Combining all components yields the SINR expression in \eqref{eQn:DLSINR}.
\end{IEEEproof}

\section{Simulation Setup and Results}

We evaluate the performance of HmMIMO against CFmMIMO and CmMIMO in terms of per-user SE over a $1\,$km$\times1\,$km coverage area.  For HmMIMO and CmMIMO this area is divided into four $500\,$m$\times500\,$m cells.  In all cases, $512$ antennas serve $32$ UEs.
In the CFmMIMO scenario, $128$ APs, each with 4 antennas, are placed uniformly at random over the entire square.  In CmMIMO, each of the four cells hosts one BS equipped with 128 co-located antennas, each BS serving 8 UEs.  In HmMIMO, per cell, $N_b=64$ antennas are centralized in a cBS, while the remaining 64 antennas are split among 16 eAPs (4 antennas each) along the cell boundaries.  This arrangement reduces the number of distributed radio heads by 50\% compared to CFmMIMO, cutting fronthaul deployment cost approximately in half.  We perform $10^5$ independent drops, randomizing AP/eAP and UE locations in each drop to capture ergodic performance.

\begin{figure}[!t]
    \centering
    \subfloat[]{
    \includegraphics[width=0.45\textwidth]{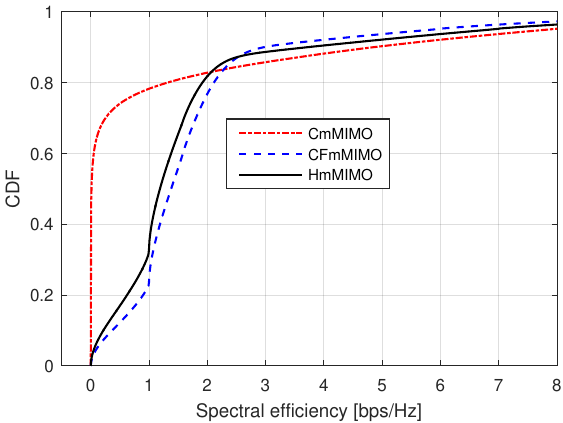} 
    \label{fig:siml}
    }         
    \caption{Performance comparison of cellular, cell-free, and heterogeneous massive MIMO networks.  }
    \label{fig:results}
\end{figure}

Thermal noise is $-174\,$dBm/Hz power spectral density with a $9\,$dB noise figure over $5\,$MHz bandwidth.  Uplink users transmit at full power ($\eta_k=1$).  All arrays are uniform linear arrays with half-wavelength element spacing; small-scale fading includes Gaussian local scattering with $15^\circ$ angular spread \cite{Ref_bjornson2020making}, while large-scale fading follows the same setting as \cite{Ref_jiang2021impactcellfree}. 
Each coherence block consists of $\tau_c=200$ channel uses (equivalent to 2ms $\times$ 100kHz), with $\tau_p=8$ uses reserved for orthogonal uplink pilots. 
\figurename~\ref{fig:results} shows the cumulative distribution functions of per-user SE.  We highlight the 95\%-likely rate (the 5th percentile of the per-user SE CDF) to gauge cell-edge performance. 
CmMIMO exhibits deep outages at the cell edges (long left tail), CFmMIMO maximizes the 95\%-likely rate—ensuring uniform service at the expense of peak rates for center UEs—and HmMIMO offers a compromise, significantly boosting the 95\%-likely rate versus CmMIMO while still delivering high peak throughput.  By using only half the distributed radio heads compared to CFmMIMO, HmMIMO achieves roughly a 50\% reduction in fronthaul cost without sacrificing macro-diversity gains.

\section{Conclusion}

We have introduced heterogeneous massive MIMO or HmMIMO, a novel cellular architecture that combines centralized large-scale antenna arrays with strategically deployed edge access points.  In each cell, a cBS hosts a high-density antenna array at its core, while multiple eAPs are distributed along cell boundaries or in coverage gaps.  The cBS orchestrates uplink and downlink processing across the eAPs via a simplified fronthaul.   Through numerical results, we have shown that HmMIMO achieves cell-edge performance on par with that of conventional cell-free systems, while substantially outperforms cellular massive MIMO, requiring far less the fronthaul infrastructure of a fully distributed cell-free deployment.  This balanced design delivers both high peak rates for central users and robust worst-case rates at the cell edge, all at a substantially reduced implementation cost.

\bibliographystyle{IEEEtran}
\bibliography{IEEEabrv,Ref_COML}

\end{document}